\documentclass[useAMS,usenatbib]{mn2e}

\RequirePackage{xspace}
\RequirePackage{amsmath}
\RequirePackage{amsfonts}
\RequirePackage{amssymb}

\def\ifundefined#1{\expandafter\ifx\csname#1\endcsname\relax}

\def\la{\mathrel{\hbox{\rlap{\hbox{\lower4pt\hbox{$\sim$}}}\hbox{$<$}}}}
\def\ga{\mathrel{\hbox{\rlap{\hbox{\lower4pt\hbox{$\sim$}}}\hbox{$>$}}}}

\newcommand{\be}{\begin{equation}}
\newcommand{\ee}{\end{equation}}

\newcommand{\bea}{\begin{eqnarray}}
\newcommand{\eea}{\end{eqnarray}}

\ifundefined{ensuremath}\def\ensuremath#1{\relax\ifmmode{#1}}
\else${#1}$\fi\else\relax\fi
\ifundefined{nuc}\def\nuc#1#2{\relax\ifmmode{}^{#1}{\protect\text{#2}}
\else${}^{#1}$#2\fi}\else\relax\fi

\newcommand{\kmps}{\ensuremath{\text{km}~\text{s}^{-1}}\xspace}

\newcommand{\msol}{\ensuremath{{\text{M}_\odot}}\xspace}

\newcommand{\nni}{\ensuremath{\nuc{56}{Ni}}\xspace}

\usepackage{graphicx}
\usepackage[T1]{fontenc} 
\usepackage{aecompl}
\usepackage{aas_macros}
\usepackage{color}

\graphicspath{{./figs/}}
\def\altaffilmark#1{$^{#1}$}
\def\altaffiltext#1#2{$^{#1}$#2}
\newcounter{aaffilcoun}
\newcommand\theaaffil{\addtocounter{aaffilcoun}{1}\theaaffilcoun}
\newcounter{affilcoun}
\newcommand\theaffil{\addtocounter{affilcoun}{1}\theaffilcoun}
\usepackage{xargs}                      
\newcommand{\mni}{\ensuremath{M_{\nuc{56}{Ni}}\xspace}}

\title[Direct Analysis of  SN 2014J]{Supernova 2014J at M82:
  II. Direct Analysis of A Middle-Class Type Ia Supernova}

\author[Vallely et al.]{Patrick Vallely\altaffilmark{\theaaffil},
  M.E. Moreno-Raya\altaffilmark{\theaaffil},
  E. Baron\altaffilmark{1,\theaaffil},
 Pilar  Ruiz-Lapuente\altaffilmark{\theaaffil,\theaaffil},
\newauthor
I. Dom{{\'i}}nguez\altaffilmark{\theaaffil}, 
Llu{{\'i}}s  Galbany\altaffilmark{\theaaffil,\theaaffil}, 
 J. I. Gonz\'alez Hern\'andez\altaffilmark{\theaaffil,\theaaffil}, 
J. M\'endez\altaffilmark{\theaaffil}, 
\newauthor
M. Hamuy\altaffilmark{8,7}, 
A. R. L\'opez-S\'anchez\altaffilmark{\theaaffil,\theaaffil}
S. Catal\'an\altaffilmark{\theaaffil}, 
E. Cooke\altaffilmark{\theaaffil},
\newauthor
C. Fari\~na\altaffilmark{11},
R. G\'enova-Santos\altaffilmark{9,10}, 
R. Karjalainen\altaffilmark{11}, 
H. Lietzen\altaffilmark{9,10}, 
J. McCormac\altaffilmark{14}, 
\newauthor
F. Riddick\altaffilmark{11}, 
J. A. Rubi\~no-Mart\'in\altaffilmark{9,10}, 
I. Skillen\altaffilmark{11}, 
\newauthor
V. Tudor\altaffilmark{11}, 
O. Vaduvescu\altaffilmark{11}\\
\altaffiltext{\theaffil}{Homer L.~Dodge Department of Physics and Astronomy,
  University of Oklahoma, 440 W.~Brooks, Rm 100, Norman, OK,
  73019-2061 USA}\\
\altaffiltext{\theaffil}{Departamento de Investigacio\'n B\'asica, CIEMAT, Avda. Complutense 40, 28040, Madrid, Spain}\\
\altaffiltext{\theaffil}{Hamburger Sternwarte, Gojenbergsweg 112, 21029
  Hamburg, Germany}\\
\altaffiltext{\theaffil}{Instituto de F{{\'i}}sica Fundamental, Consejo
  Superior de Investigaciones Cient\'ficas, c/. Serrano 121, E-28006,
  Madrid, Spain} \\
\altaffiltext{\theaffil}{Institut de Ci{\'e}ncies del Cosmos (UB-IEEC),
  c/. Mart\'i i i Franqu{\'e}s 1, E-08028, Barcelona, Spain} \\
\altaffiltext{\theaffil}{Universidad de Granada, E-18071, Granada, Spain} \\
\altaffiltext{\theaffil}{Millennium Institute of Astrophysics, Universidad de Chile, Chile}\\
\altaffiltext{\theaffil}{Departamento de Astronom\'ia, Universidad de
  Chile, Santiago, Chile} \\
\altaffiltext{\theaffil}{Instituto de Astrof\'isica de Canarias,
  38200 La Laguna, Tenerife, Spain} \\
\altaffiltext{\theaffil}{Departmento de Astrof\'isica, Universidad de
  La Laguna (ULL),
  38206 La Laguna, Tenerife, Spain} \\
\altaffiltext{\theaffil}{Isaac Newton Group of Telescopes, Apto. 321,
  E-38700 Santa Cruz de la Palma, Canary Islands, Spain} \\
\altaffiltext{\theaffil}{Australian Astronomical Observatory, PO BOX
  296, Epping, NSW 1710, Australia} \\
\altaffiltext{\theaffil}{Department of Physics and Astronomy,
  Macquarie University, NSW 2109, Australia} \\
\altaffiltext{\theaffil}{Department of Physics, University of Warwick,
  Coventry CV47AL, UK} \\
\altaffiltext{\theaffil}{School of Physics and Astronomy, University
  of Nottingham, University Park, Nottingham, NG7 2RD, UK} \\
}

\begin{document}

\date{Accepted xxx Received xx; in original form xxx}

\pagerange{\pageref{firstpage}--\pageref{lastpage}} \pubyear{2015}

\maketitle

\label{firstpage}

\begin{abstract}
We analyze a time series of optical  spectra of SN~2014J from
almost two weeks prior to maximum to nearly four months after
maximum. We perform our analysis using the SYNOW code, which is
well suited to track the distribution of the ions
with velocity in the ejecta.
We show that almost all of the spectral features during the
entire epoch can be identified with permitted transitions of the
common ions found in normal SNe~Ia in agreement with
previous studies.

We show that 2014J is a relatively normal SN Ia. At early times the
spectral features are dominated by Si II, S II, Mg II, and Ca
II. These ions persist to maximum light with the appearance of Na I
and Mg I. At later times iron-group elements also appear, as expected
in the stratified abundance model of the formation of normal type Ia
SNe. 

We do not find significant spectroscopic evidence for oxygen, until
100 days after maximum light. The +100 day identification of oxygen is
tentative, and 
would imply significant mixing of unburned or only slight processed
elements down to a velocity of 6,000~\kmps.
Our results are in relatively good agreement with other analyses in
the IR. We briefly compare SN~2011fe to SN~2014J and conclude that the
differences could be due to different central densities at ignition or
differences in the C/O ratio of the progenitors.
\end{abstract}

\section{Introduction}

SN 2014J, located in M82 only a 3.3  Mpc away \citep{Foley14J14}, is the closest
SN Ia observed in the last 40 years. 
As the nearest modern SN~Ia, SN~2014J has been extremely well
observed: in $\gamma$-rays \citep{Diehl14J14,churazov14J14}, X-rays
\citep{Margutti14J14}, by HST
\citep{Foley14J14,kelly14J14}, optical
\citep{Goobar14j14,Nielsen14J14,amanullah14J14,ashall14J14} at high resolution
  \citep{Welty14J14,ritchey14J14,jack14J15,Graham14J15}, near-IR \citep{marion14J15,brian_14JIR14}, mid-IR \citep{Johansson14J15,telesco14J15},
polarimetry \citep{Kawabata14J14}, with rapid photometry
\citep{Siverd14J15}, and in radio \citep{Torres14J14}. 

In \citet{galbany14J15} (henceforth Paper I) 
we present a long time baseline set of optical spectroscopy and we
analyzed the dataset using wavelength matching to obtain line
identifications and velocities. By the term wavelength matching
we mean that no modeling of the spectra was performed, the line
IDs were simply made by matching the wavelength of an observed feature
to known wavelengths of characteristic lines. Here we take that
analysis one step 
further using the 
parameterized synthetic-spectrum code SYNOW \citep{jb90,bbj03}. In the
SYNOW framework,  
we model the  observed 
spectra, identify the line features, and determine the photospheric
velocity and the velocity extent of the ions  of SN~2014J at each epoch. 

The SYNOW code is based on simple assumptions, described in more
detail below.
A synthetic spectrum consists of blended P Cygni profiles (emission
component,  blueshifted  absorption  component) 
superimposed  on  a  continuum.  SYNOW  does  not  do 
continuum  transport,  it  does  not  solve  rate  equations,  and  it
does not calculate ionization ratios. Its main function is to take
multiple line scattering into account so that it can be used in
an empirical spirit to make line identifications and estimate the
velocity  at  the  photosphere  (or  pseudophotosphere),  and  the
velocity interval within which each ion is detected. These quantities
provide  constraints  on  the  composition  structure  of  the 
ejected matter.

As described above, SYNOW is a parameterized line scattering code that includes the
effects for multiple scattering. An opaque photosphere illuminates 
the atmosphere with a blackbody. The velocity is homologous, $v
\propto r$, and the photosphere is at a velocity coordinate
$v_\text{ph}$. The source function is taken to be that of resonance
scattering $S = W(r) I_\text{ph}$, where $I_\text{ph}$ is the intensity
from the photosphere and 
\[ W = \frac{1}{2}\left\{1 - \sqrt{1-(\frac{r_\text{ph}}{r})^2}\right\} \]
is the dilution factor \citep{mihalas78sa}, and $r_\text{ph}$ is the
photospheric radius. 

Lines are treated in the Sobolev approximation. The Sobolev optical
depth, $\tau_\text{sob}$,  is a local quantity, given by
\[ \tau_\text{sob} = \left(\frac{\pi e^2}{m_e c}\right) f_0 n_l (1-\frac{g_l n_u}{g_u
  n_l}) \lambda_0/\frac{dv}{dr}, \] where $l$ designates the lower
level of the transition, $u$ the upper level, $f_o$ is the oscillator
strength of the line, $n_l$ and $n_u$ are the occupation numbers of
the levels and $g_l$ and $g_u$ are the statistical weights of the
levels \citep{jb90}. 

In SYNOW, a reference line is chosen for each ion and $n_l$ is assumed
to vary as $e^{-v/v_e}$. $v_e$ is the density scale factor in units of \kmps and its value
determines the velocity extent of the absorption trough. Thus, $\tau$
for the reference line is given 
as $\tau(v) = \tau e^{-v/v_e}$, where $\tau$ is the value of the
Sobolev optical depth where the ion begins, usually at the photosphere,
$v_\text{ph}$, but the line may be detached and begin at a velocity
$v_{\text{min}_e}$. Everything else about the ion is 
assumed to be given by the Boltzmann formulae with a ion temperature
$T_\text{exe}$. 
For more information about the SYNOW code see \citet{jb90} and 
\citeauthor{branch00cx04}~(\citeyear{branch00cx04}; \citeyear{branchcomp105};
\citeyear{branch_pre07}).
\nocite{branch00cx04,branchcomp105,branch_pre07} The specific
  values optimized by eye
are $v_\text{ph}$, $T_{\text{bb}}$, $T_{\text{exe}}$, and the
velocity and optical depth of each ion.
Values for each input parameter are determined by manually varying
them and examining the resultant spectra until the best possible
match by eye is obtained. 
These parameters are determined by the quality of
the model. $v_\text{ph}$ is determined by global constraints across
all lineshapes. 
$v_{\text{min}_e}$  is
used only if detachment is deemed to be required after 
first trying hard without detached lines. $T_\text{exe}$ is not a terribly
important parameter and its value is pretty much kept at the default
value unless there is a need to have more or less abundance in upper
or lower states of a given ion. The error estimates for $\tau$ were
obtained by varying the 
parameter values and choosing an acceptable range as judged by eye.

While this by eye pattern matching is standard in the community
  \citep[see, for example,][and references therein]{jerod11fe12}, one
  could go further in terms of fitting and try to  
construct some form of complex likelihood combined
with Markov Chain Monte Carlo, it would be overkill for the intended
purpose of ion identification and velocity extent that is the goal of
this and similar analyses. To our knowledge it has never been done and
is beyond the scope of the present work.

The spectral epochs range
from almost two weeks pre-maximum to nearly three months post-maximum
with wavelengths from roughly 3500 {\AA} to 9500 {\AA}. We take the
time of maximum to be February 1, 2014 \citep{galbany14J15}. The spectra
are corrected to account for being heavily reddened by dust in
the host galaxy  as well as for foreground
extinction. We use the reddening values found by
\citet{Goobar14j14}. We use this multi-epoch spectral dataset
  to identify the ions in the spectra as a function of time and
  analyze their velocity extent. Since Type Ia supernovae play such an
  important role as cosmological probes, their progenitor system is uncertain
  and the explosion mechanism is not well understood \citep[for recent
  reviews see][]{hill_rev13,MMN14,pilar_rev14}, detailed
  analyses of the explosion products of well-observed SNe Ia are
  particularly useful to gain insight into these unresolved questions.
A good introduction to the
analysis of supernova spectra is given in 
\citet[][and references therein]{parrent14}. In \S~\ref{sec:premax} we
identify the ions and their velocity extent for epochs up to Day
-4.6. In \S~\ref{sec:nearmax} we identify the ions and their velocity
extent for epochs near maximum light, up to Day
+12.4. \S~\ref{sec:postmax} studies the epochs up until
Day 
+22.5,  and \S~\ref{sec:late} examines late epochs between Day +33.4
and +99.1. In \S~\ref{sec:vphotcompare} we discuss the velocities we
have obtained for the photosphere and compare those to what others
have found for SN~2014J and other Branch normal
supernovae. \S~\ref{sec:compare} compares the results we have found
for SN~2014J directly with those of SN~2011fe and describes possible
reasons for similarities and differences between the two
supernovae. Finally, in \S~\ref{sec:conclusion} we recap and present
our conclusions.

\section{Pre-Maximum Spectra}
\label{sec:premax}
Observational details of all of the data used for this work can be found in Paper I. Our earliest observed spectrum was obtained on January 23 (Day -9.6), and we examine a  total of 6 spectra obtained between
January 23 (Day -9.6) and January 28 (Day -4.6).The pre-maximum spectra all show
standard SNe~Ia features: Si~II $\lambda 6355$, Si~II $\lambda 5972$
S~II
absorption centered at 5300 {\AA}, Mg II absorption centered at 4400
{\AA} as well as 7600 {\AA} (Mg II $\lambda\lambda 7889.9$), Mg I $\lambda\lambda 5178.3$, and Ca II H+K (absorption centered around
3700 {\AA}) and the IR triplet with absorption centered 
at 8200 {\AA}. In this time period there is only minimal evolution. The
SYNOW spectra 
include lines of 5 ions: Mg I, Mg II, Si II, S II, Ca II. In
addition to the photospheric component, Ca II has a detached
high-velocity component, and a detachment velocity of $v_{{min}_e}=23,000$ \kmps
is used for
all of the pre-maximum models. Values of $v_{\text{phot}}=14,000$ \kmps,
and $T_{\text{exe}}=10,000$ K are used for all ions. A complete list of input
parameters can be found in Table~\ref{tab:premaxparams}.

\begin{figure}
\centering
\includegraphics[scale=0.55]{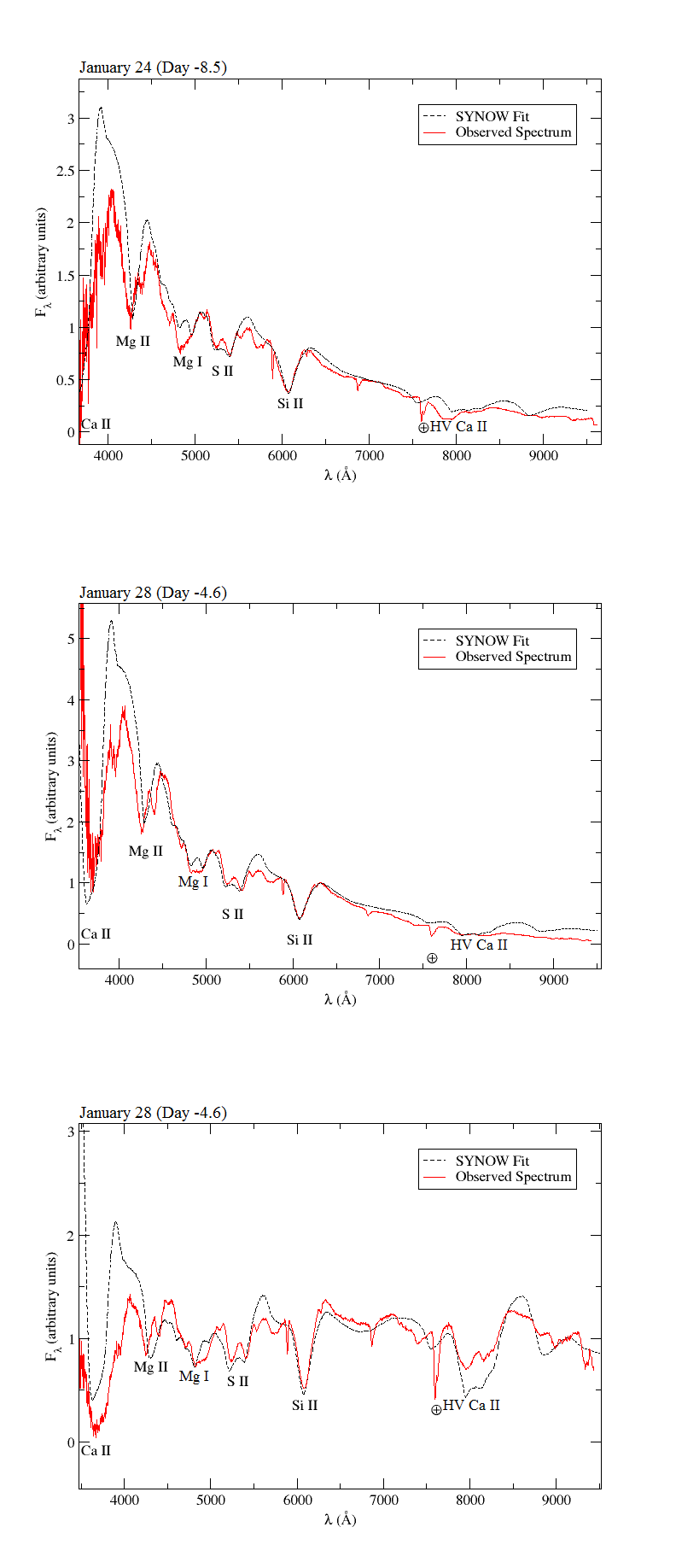}
\caption{\label{fig:premax}The observed spectra and SYNOW model representative of the pre-maximum epoch. The upper panel shows the January 24 (Day -8.5) spectra, the middle panel shows the January 28 (Day -4.6) spectra, and the bottom panel shows the flattened January 28 (Day -4.6) spectra.}
\end{figure}

The upper panel of Figure~\ref{fig:premax} shows the observed spectrum obtained on January
24 (Day -8.5) compared with the SYNOW model. Here, high
velocity Ca is modelled 
with a detached high-velocity component.

The middle panel of Figure~\ref{fig:premax} shows the observed spectrum obtained on January
28 (Day -4.6) and the SYNOW model, and the lower panel shows
the flattened spectra, where the spectra have been normalized by the
local (pseudo) continuum in that wavelength region using the
prescription of \citet{jeffery07a}. The Ca II detached high-velocity
component persists through this epoch. Flattened spectra are
  often presented in the 
literature, so we include them for the benefit of readers who are used
to seeing them. The models were calculated on the original,
unflattened spectra.

The photospheric velocity remains constant throughout this period, the
Mg~II lines strength is constant (that is, $\tau$ doesn't change), Si~II strengthens, S~II strengthens,
photospheric Ca~II lines get stronger, and the high velocity feature
is present throughout. While Ca H+K is at the blue edge of these
spectra, it is a distinctive and relatively unblended feature and the
Ca II IR triplet also provides a constraint.

We do not find strong spectroscopic evidence for O~I $\lambda 7774$,
as O~I is not used in the SYNOW model,
whereas abundance tomography finds that the high velocity ejecta is
dominated by oxygen \citep{ashall14J14}. We will discuss this
discrepancy in more detail in \S~\ref{sec:conclusion}.

\begin{table*}
\begin{tabular}{ccccccccccccccccc}
\multicolumn{11}{c}{Pre-Maximum SYNOW Parameters}\\
\hline\hline
Date&
$v_{\text{phot}}$&
$T_{\text{bb}}$&
$T_{\text{exe}}$&
  \multicolumn{2}{c}{Mg I} & \multicolumn{2}{c}{Mg II} & \multicolumn{2}{c}{Si II} & \multicolumn{2}{c}{S II} & \multicolumn{2}{c}{Ca II} & \multicolumn{3}{c}{HV Ca II} \\
&
&
&
&
$\tau$ &
$v_e$ &
$\tau$ & 
$v_e$ & 
$\tau$ & 
$v_e$ & 
$\tau$ & 
$v_e$ & 
$\tau$ &
$v_e$ &
$\tau$ &
$v_e$ & 
$v_{\text{min}_e}$\\
\hline\hline
 Jan 23(-9.6) & 14 & 14 & 10 & 1$\pm$0.1 & 2 & 2.5$\pm$0.5 & 2 & 2$\pm$0.2 & 3 & 1.3$\pm$0.1 & 2 & 15$\pm$2.5 & 3 & 10$\pm$2 & 3 & 23 \\
 Jan 24(-8.5) & 14 & 14 & 10 & 1$\pm$0.1 & 2 & 2.5$\pm$0.5 & 2 & 2$\pm$0.2 & 3 & 1.3$\pm$0.1 & 2 & 15$\pm$2.5 & 3 & 10$\pm$2 & 3 & 23 \\ 
 Jan 25(-7.4) & 14 & 17 & 10 & 0.8$\pm$0.1 & 2 & 2$\pm$0.4 & 2 & 3$\pm$0.3 & 2 & 1$\pm$0.1 & 2 & 20$\pm$4 & 3 & 10$\pm$2 & 3 & 23 \\
 Jan 26(-6.5) & 14 & 18 & 10 & 1$\pm$0.1 & 2 & 2$\pm$0.4 & 2 & 3$\pm$0.3 & 2 & 1$\pm$0.1 & 2 & 25$\pm$5 & 3 & 10$\pm$2 & 3 & 23 \\
 Jan 27(-5.4) & 14 & 22 & 10 & 1$\pm$0.1 & 2 & 2$\pm$0.4 & 2 & 4$\pm$0.4 & 2 & 1.2$\pm$0.1 & 2 & 25$\pm$5 & 3 & 10$\pm$2 & 3 & 23 \\
 Jan 28(-4.6) & 14 & 24 & 10 & 1$\pm$0.1 & 2 & 2$\pm$0.4 & 2 & 4$\pm$0.4 & 2 & 1.7$\pm$0.2 & 2 & 35$\pm$7 & 3 & 20$\pm$3 & 3 & 23 \\
\hline\hline
\end{tabular}
\caption{SYNOW parameters for Pre-Maximum spectra model. \label{tab:premaxparams}}
\end{table*}

\section{Near-Maximum Spectra}
\label{sec:nearmax}
Maximum light occurred on February 1, 2014, and we examine
 six spectra obtained
between February 4 (Day +2.3) and February 14 (Day +12.4). 
SYNOW spectra include lines of the following ions: Na I,
Mg I, Mg II, Si II, Si III, S II, and Ca II. 
A complete list of input parameters
is given in Table~\ref{tab:nearmaxparams}. Values of
$v_{\text{phot}}=12,000$ \kmps, and $T_{e}=10,000$ K are used for all ions
except for the February 14 (Day +12.4) spectra, in which a value of
$v_{\text{phot}}=11,000$ \kmps is used. The line identifications remain
the same as in \S~\ref{sec:premax} with the addition 
of Na ID (with absorption centered at
5700 {\AA}).
The S II
absorption feature declines steadily with 
time. In normal SNe Ia the S II W feature is
distinct and relatively easy to identify.  The detached
high-velocity Ca II component is no longer present. For spectra
  that don't cover Ca II, it was not reported in the Table. The increase
  in strength of Ca II, indicates the ejecta are cooling and Ca II
  becomes strong.
The Na I
line is likely, but the large $\tau$ found for the February 6 (Day
+4.2) spectrum 
is due to the dichroic break in the observed spectrum.   Comparing with
\citet{ashall14J14} our results are somewhat different, by this time
they find a significant amount of Fe~II, which they take as evidence
for \nni above the photosphere, we do not find strong evidence for
this. Na D are typically identified on core normal (or close to core
normal) SNe~Ia about one week post maximum light \citep{branch_post08}.

\begin{figure*}
\centering
\includegraphics[scale=0.55]{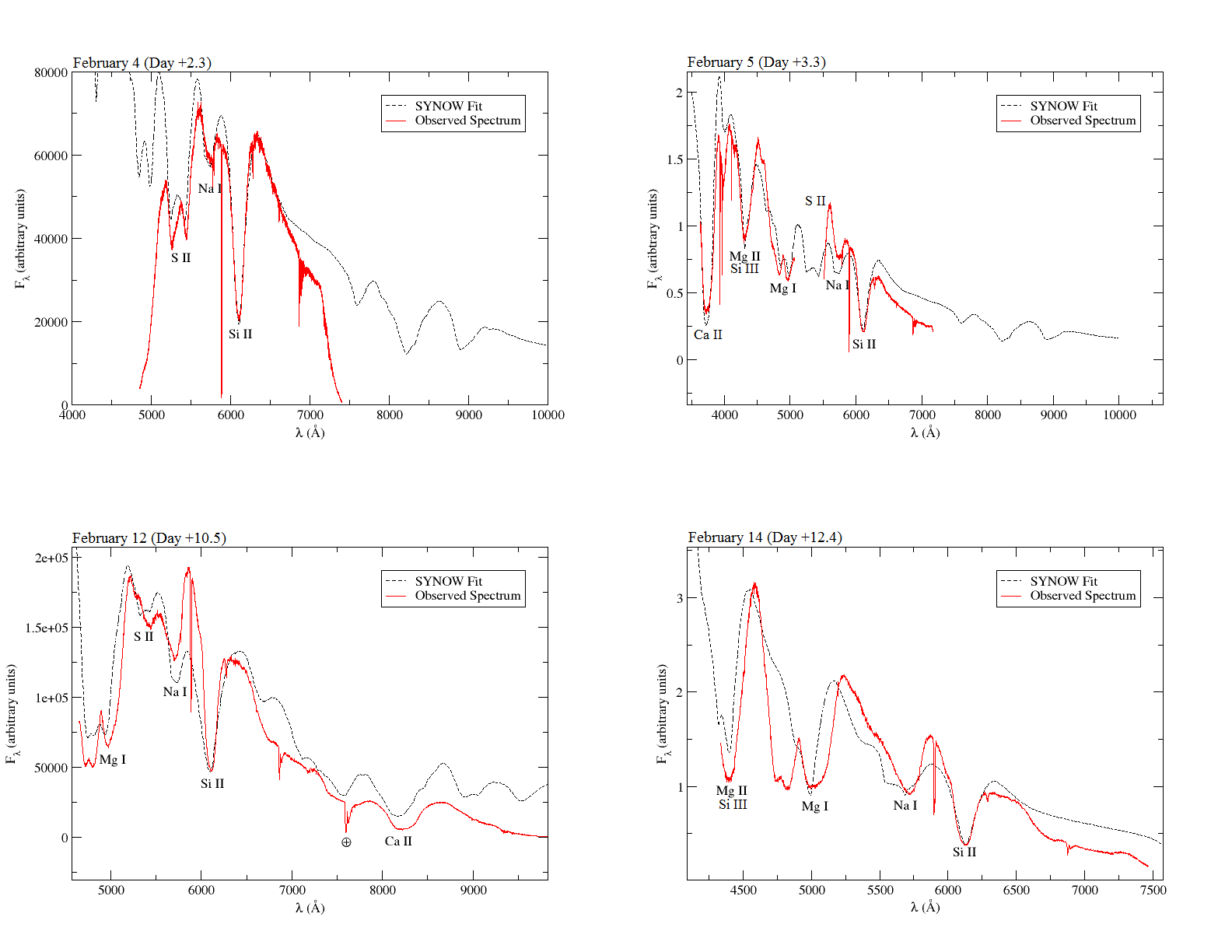}
\caption{\label{fig:nearmax}The observed spectra and SYNOW models representative of the near-maximum epoch. The upper left panel shows the February 4 (Day +2.3) spectra, the upper right panel shows the February 5 (Day +3.3) spectra, the lower left panel shows the February 12 (Day +10.5) spectra, and the lower right panel shows the February 14 (Day +12.4) spectra.}
\end{figure*}

The upper left panel of Figure~\ref{fig:nearmax} shows the observed spectrum obtained on February
4 (Day +2.3) and the SYNOW model. The blue and red arms of the observed spectrum
were matched at 6000 {\AA}. The match was simply to remove the red (blue) wing of the blue (red)
spectra due to the low transmission at those wavelengths. There was no
disagreement in the flux calibrations, but in this way we could treat
both spectra as if it were a single spectrum. Note that this is
clearly shown in Figure 3 of Paper I, e.g. spectra at epoch -7.4d was
matched at 7300~{\AA}, and spectra at epoch +2.3d at 6000~{\AA}. In the first
case (-7.4d) the two spectra have different resolutions (3.30 and 0.49
$\AA$~pix$^{-1}$), while in the second (+2.3d) they had the same
resolution (0.53 \AA~pix$^{-1}$)
and the effect is clearly not visible.

 The upper right
panel of Figure~\ref{fig:nearmax} shows the observed spectrum obtained on February 5 (Day +3.3)
and the SYNOW model. A value of $v_{\text{phot}}=12,000$ \kmps is used. Si III is now used in the model,
to help with the absorption produced by the Si~III
$\lambda\lambda 4425.3$ line and the absorption in the Na I D trough
produced by the Si~III $\lambda\lambda 5707.5$ line. Since it is only
used to help with absorption in blended features, the Si~III
identification is tentative.  

The lower left panel of Figure~\ref{fig:nearmax} shows the observed spectrum obtained on February
12 (Day +10.5) and the SYNOW model. It exhibits
Mg~I absorption centered at 5000 {\AA} and Mg II absorption
centered at 7600 {\AA} and 8900 {\AA}.  A value of $v_{\text{phot}}=12,000$ 
\kmps is used. 

The lower right panel shows the observed spectrum obtained on February
14 (Day +12.4) and the SYNOW model.  This is nearly identical
to the Day +10.5 model, but with Si~III, and weaker Si II.

Looking at the four panels in Figure~\ref{fig:nearmax}, it appears as
if the model of the Mg~I doublet degrades with epoch from February 4 to
12, however, in reality the two features seen in absorption at
4750~\AA\ and 5000~\AA\ are stronger in the last epoch and thus,
the Mg~I identification on day +12.4 is
provisionary. That is, while we 
  believe that the feature is produced to some extent by Mg I, without
  determining the other ions that produce the blended feature we can
  not be sure of the velocity extent of Mg I.

\begin{table*}
\begin{tabular}{cccccccccccccccccc} 
\multicolumn{12}{c}{Near-Maximum SYNOW Parameters}\\
\hline\hline
Date&
$v_{\text{phot}}$&
$T_{\text{bb}}$&
$T_{\text{exe}}$&
 \multicolumn{2}{c}{Na I} & \multicolumn{2}{c}{Mg I} & \multicolumn{2}{c}{Mg II} & \multicolumn{2}{c}{Si II} & \multicolumn{2}{c}{Si III} & \multicolumn{2}{c}{S II} & \multicolumn{2}{c}{Ca II} \\
&
&
&
&
$\tau$ &
$v_e$ &
$\tau$ & 
$v_e$ & 
$\tau$ & 
$v_e$ & 
$\tau$ & 
$v_e$ & 
$\tau$ &
$v_e$ &
$\tau$ &
$v_e$ & 
$\tau$ &
$v_e$\\
\hline\hline
 Feb 04(+2.3) & 12 & 11 & 10 & 1$\pm$0.1 & 2 & --- & --- & 2$\pm$0.4 & 2 & 6$\pm$0.6 & 2 & 1$\pm$0.3 & 1 & 2.2$\pm$0.2 & 2 & --- & --- \\
 Feb 05(+3.3) & 12 & 11 & 10 & 1$\pm$0.1 & 2 & 2$\pm$0.2 & 2 & 2$\pm$0.4 & 2 & 6$\pm$0.6 & 2 & 1$\pm$0.3 & 1 & 1.2$\pm$0.1 & 2 & 30$\pm$6 & 3 \\
 Feb 06(+4.2) & 12 & 11 & 10 & 10$\pm$1 & 2 & 5$\pm$0.5 & 2 & 2$\pm$0.4 & 2 & 6$\pm$0.6 & 2 & 1$\pm$0.3 & 1 & 1.2$\pm$0.1 & 2 & --- & --- \\
 Feb 12(+10.5) & 12 & 11 & 10 & 1$\pm$0.1 & 2 & 6$\pm$0.5 & 2 & 0 & 2 & 6$\pm$0.6 & 2 & 0 & 1 & 0.3$\pm$0.1 & 2 & 180$\pm$40 & 3 \\
 Feb 14(+12.4) & 11 & 22 & 10 & 1$\pm$0.1 & 2 & 3$\pm$0.3 & 2 & 2$\pm$0.4 & 2 & 4$\pm$0.4 & 2 & 10$\pm$3 & 1 & 0 & 2 & --- & --- \\
\hline\hline
\end{tabular}
\caption{SYNOW Parameters for Near-Maximum spectra models.
\label{tab:nearmaxparams}}
\end{table*}

\section{Post-Maximum Spectra}
\label{sec:postmax}
Six spectra were obtained between February 19 (Day +17.4) and February 26 (Day +24.5). The
post-maximum spectra share the same general features --- Si II
absorption centered at 6100 
{\AA}, Mg II absorption centered at 7600 {\AA}, Ca II absorption
centered at 8200 {\AA}, Fe II absorption centered at 8900 {\AA}, and
Na I and Si III 
absorption centered at 5600 {\AA}, with some variation. 
Since the Si~III is there just to help in the Na I
D trough, and it is physically unlikely to have such highly excited
ions at this stage, we regard the Si~III as weak at this epoch. The
SYNOW spectra include lines of 7 ions: Na 
I, Mg I, Mg II, Si II, Si III, Ca II, and Fe II. Values of
$v_{\text{phot}}=11,000$ \kmps and  $T_{e}=10,000$ K are used for all ions
except for the February 24 (Day +22.5) and February 26 (Day +24.5)
spectra, where  
$v_{\text{phot}}$ is reduced to $10,000$ \kmps. 
Table~\ref{tab:postmaxparams} lists the input parameters. 

\begin{figure*}
\centering
\includegraphics[scale=.55]{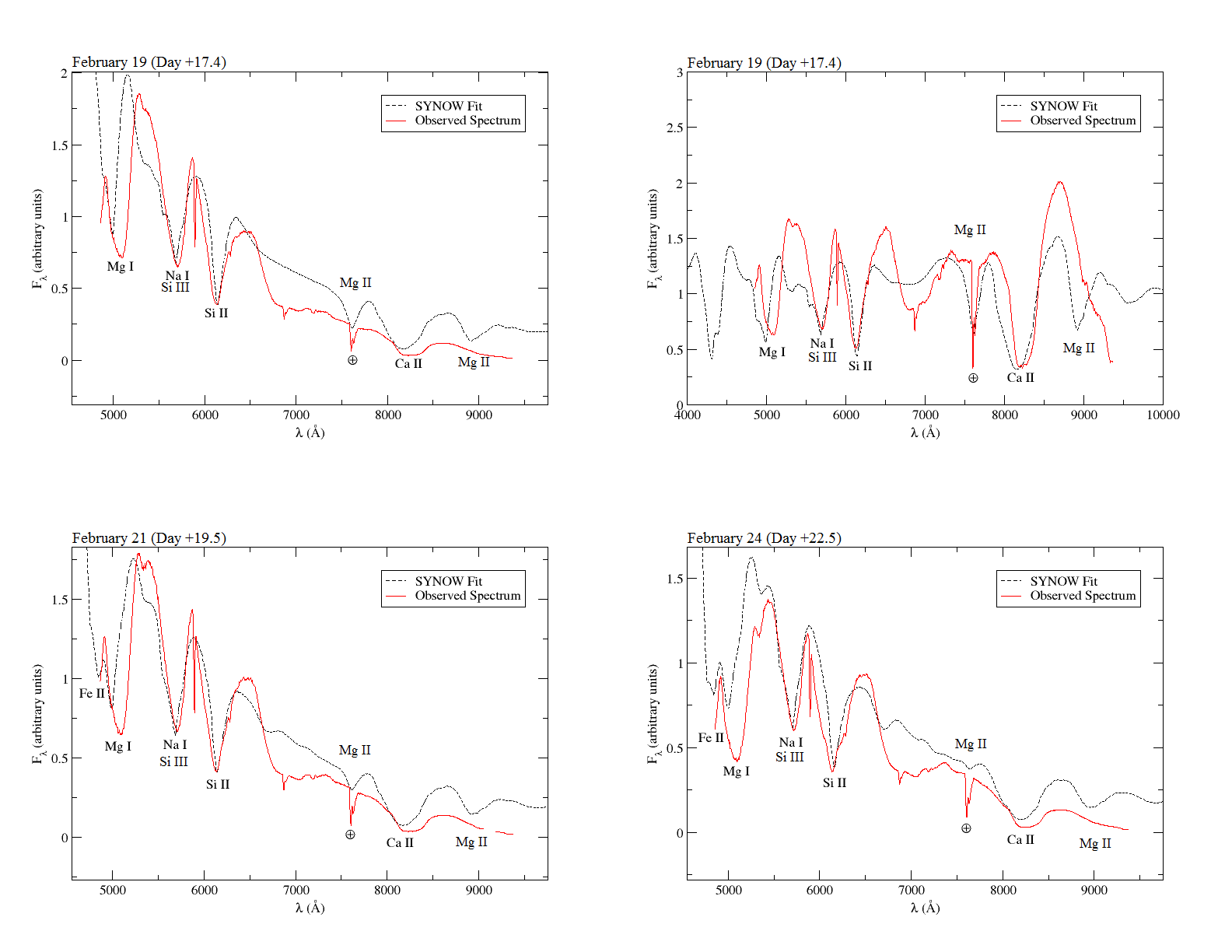}
\caption{\label{fig:postmax}The observed spectra and SYNOW models representative of the post-maximum epoch. The upper left panel shows the February 19 (Day +17.4) spectra, the upper right panel shows the flattened February 19 (Day +17.4) spectra, the lower left panel shows the February 21 (Day +19.5) spectra, and the lower right panel shows the February 24 (Day +22.5) spectra.}
\end{figure*}

The upper left panel of Figure~\ref{fig:postmax} shows the observed spectrum obtained on February 19 (Day +17.4) and the SYNOW model, and the upper right panel shows the flattened spectra.
The synthetic spectrum is a decent reprodution of the
  observatons. It has several 
prominent features: Si II absorption centered at 6100 {\AA}, Na I
absorption centered at 5700 {\AA}, Mg I absorption centered at 5000
{\AA}, Mg II absorption centered at 7600 {\AA} and 8900 {\AA}, and Ca
II absorption centered at 8200 {\AA}.  A
value of $v_{\text{phot}}=11,000$ \kmps is used. 

The lower left panel of Figure~\ref{fig:postmax} shows the observed spectrum obtained on February
21 (Day +19.5) and the SYNOW model. 
This epoch marks the beginning of the Fe II presence that characterizes
the later synthetic spectra with a value of $\tau=3.0$ and the
continued presence of the blackbody bump around 7000 {\AA}. A value of
$v_{\text{phot}}=11,000$ \kmps is used. 

The lower right panel shows the observed spectrum obtained on February
24  (Day +22.5) and the SYNOW model. As is characteristic of the Post-Maximum
Spectra, the synthetic spectrum is  decent. Fe~II is growing in
strength to 
$\tau=6.0$  A value of
$v_{\text{phot}}=10,000$ \kmps is used. 

\begin{table*}
\begin{tabular}{cccccccccccccccccc} 
\multicolumn{12}{c}{Post-Maximum SYNOW Parameters}\\
\hline\hline
Date&
$v_{\text{phot}}$&
$T_{\text{bb}}$&
$T_{\text{exe}}$&
 \multicolumn{2}{c}{Na I} & \multicolumn{2}{c}{Mg I} & \multicolumn{2}{c}{Mg II} & \multicolumn{2}{c}{Si II} & \multicolumn{2}{c}{Si III} & \multicolumn{2}{c}{Ca II} & \multicolumn{2}{c}{Fe II} \\
&
&
&
&
$\tau$ &
$v_e$ &
$\tau$ & 
$v_e$ & 
$\tau$ & 
$v_e$ & 
$\tau$ & 
$v_e$ &
$\tau$ &
$v_e$ & 
$\tau$ &
$v_e$ & 
$\tau$ &
$v_e$\\
\hline\hline
 Feb 19(+17.4) & 11 & 22 & 10 & 2.5$\pm$0.3 & 2 & 3$\pm$0.3 & 2 & 6$\pm$1.2 & 2 & 12$\pm$1.2 & 1 & 10$\pm$3 & 1 & 200$\pm$50 & 3 & 0 & 2 \\
 Feb 20(+18.3) & 11 & 22 & 10 & 2.5$\pm$0.3 & 2 & 3$\pm$0.3 & 2 & 4$\pm$0.8 & 2 & 10$\pm$1 & 1 & 10$\pm$3 & 1 & 180$\pm$40 & 3 & 0 & 2 \\
 Feb 21(+19.5) & 11 & 22 & 10 & 2.5$\pm$0.3 & 2 & 3$\pm$0.3 & 2 & 3$\pm$0.6 & 2 & 10$\pm$1 & 1 & 10$\pm$3 & 1 & 180$\pm$40 & 3 & 3$\pm$0.8 & 2 \\
 Feb 24(+22.5) & 10 & 22 & 10 & 2.5$\pm$0.3 & 2 & 3$\pm$0.3 & 2 & 1$\pm$0.2 & 2 & 10$\pm$1 & 1 & 10$\pm$3 & 1 & 140$\pm$35 & 3 & 6$\pm$1.5 & 2 \\
 Feb 26(+24.5) & 10 & 22 & 10 & 2$\pm$0.2 & 2 & 3$\pm$0.3 & 2 & 2$\pm$0.4 & 2 & 12$\pm$1.2 & 1 & 14$\pm$4 & 1 & 140$\pm$35 & 3 & 6$\pm$1.5 & 2 \\
\hline\hline
\end{tabular}
\caption{SYNOW Parameters for Post-Maximum spectra. \label{tab:postmaxparams}
}
\end{table*}

\section{Late Spectra}
\label{sec:late}
The final spectrum we examined was obtained on May 11 (Day +99.1), and
we have 8 spectra obtained between March 6 (Day +33.4) and May 11 (Day
+99.1). The late spectra share some general features --- Na I
absorption centered at 5600 {\AA}, Mg II absorption centered at 7600
{\AA}, Ca II absorption centered at 8200 {\AA}, Co II absorption
centered at 5000 {\AA} and 6400 {\AA}, and Fe II absorption centered
at 8900 {\AA} --- although with far more variation than observed in
the earlier epochs. Their SYNOW spectra include lines of 7 ions: Na I,
Mg I, Mg II, Ca II, Fe II, Co II, and Ni II. The $v_{\text{phot}}$
values used decrease from 9,000 \kmps to 6,000 \kmps as the spectra
age, and $T_{\text{exe}}=6,000$ K is used for the four latest
spectra. A complete list of input parameters can be found in
Table~\ref{tab:latemaxparams}.

These later spectra where we see much deeper into the ejecta are also
likely indicating that our exponential density parameterization is too
simplistic for the entire ejecta.  Our models are very smooth, whereas
the observations show more complex line shapes. The ejecta density
profile is possibly more 
complex, which may lead to the observed line shapes, but to introduce more
complex density parameterizations is beyond the scope of this work.

These late times are not strictly in keeping with the
assumptions made in SYNOW. However, because past
Day +100 permitted lines dominate the spectra of SNe~Ia
\citep{branch_post08,brian_14JIR14}, the SYNOW-like assumptions are
not invalid at these epochs, although we do make allowance for the
variation of the ``photosphere'' with wavelength at these late times
by modeling the blue and red regimes of the spectra with different $v_{phot}$ values.
We note that SYNOW-like assumptions lead to qualitative insights even
at very late times \citep{nebsyn12}. With that in mind, we go
further in our analysis by examining spectra from these late epochs.

\begin{figure*}
\centering
\includegraphics[scale=.55]{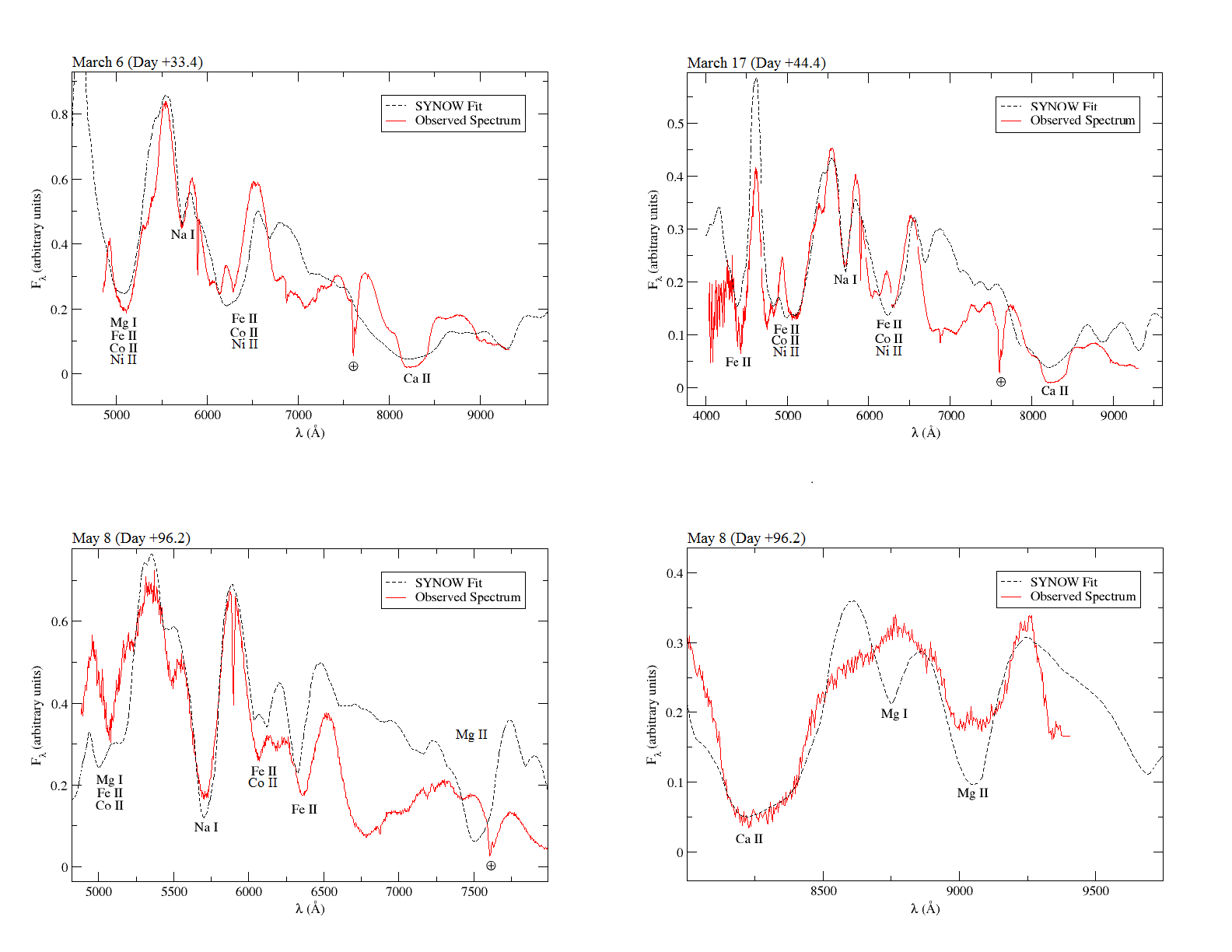}
\caption{\label{fig:late}The observed spectra and SYNOW models representative of the late epoch. The upper left panel shows the March 6 (Day +33.4) spectra, the upper right panel shows the March 17 (Day +44.4) spectra, and the lower left and lower right panels show the May 8 (Day +96.2) spectra, split into two wavelength regimes.}
\end{figure*}

The upper left panel of Figure~\ref{fig:late} shows the observed spectrum obtained on March 6 (Day +33.4)
and the SYNOW model. The SYNOW model is now strongly
affected by the 
underlying blackbody and lack of continuum transfer. 
It has several prominent features --- Na ID absorption
centered at 5700 {\AA}, Mg I absorption centered at 5000 {\AA}, Fe II absorption
centered at 5000 {\AA} and 6400 {\AA}, and Ca II absorption centered
at 8200 {\AA}. The SYNOW spectrum includes lines of 6 ions: Na I, Mg
I, Ca II, Fe II, Co II, and Ni II.
A value of $v_{\text{phot}}=9,000$ \kmps is used. A complete list
of input parameters can be found in Table~\ref{tab:latemaxparams}.

The upper right panel of Figure~\ref{fig:late} shows the observed spectrum obtained on March 17 (Day +44.4)
and the SYNOW model. It has several prominent features {\textemdash} Na I absorption centered at 5700
{\AA}, Fe II and Co II
absorption centered at 5000 {\AA} and 6400 {\AA}, and Ca II absorption
centered at 8200 {\AA}. The SYNOW spectrum includes lines of 5 ions:
Na I, Ca II, Fe II, Co II, and Ni II.

The lower left and lower right panels of Figure~\ref{fig:late} show the observed
spectrum obtained on May 8 (Day +96.2) and the SYNOW model over two wavelength
regimes. $T_{e}=6,000$~K  is used.
 Note that the two figures
are split at 8000 {\AA}. For the left hand figure,
$v_{\text{phot}}=9,000$ \kmps is 
used, whereas for the right hand figure a much slower
$v_{\text{phot}}$ of 2,000 \kmps is used, and $v_e=4$ is used for Ca II;
otherwise it uses the same parameters as that shown on the left. 
Our use of a blackbody as the inner boundary condition, coupled with
the fact that SYNOW does no continuum transfer and is a pure
Schuster-Schwarzschild method leads to a bump in the pseudo continuum
or line features just to the blue of 7000 {\AA}, this feature becomes
even more pronounced at later epochs.
The blue model continues to show Na D
absorption centered at 5700 {\AA}, Mg II absorption centered at 7600
{\AA} and 8900 {\AA}, Fe II absorption centered at 5000 {\AA} and 6400
{\AA}, and Ca II absorption centered at 8200 {\AA}. The SYNOW spectrum
includes lines of 6 ions: Na I, Mg I, Mg II, Ca II, Fe II, Co II, Ni
II.

\begin{table*}
\begin{tabular}{cccccccccccccccccccccccc}
\multicolumn{16}{c}{Late SYNOW Parameters}\\
\hline\hline
Date&
$v_{\text{phot}}$&
$T_{\text{bb}}$&
$T_{\text{exe}}$&
 \multicolumn{2}{c}{Na I} & \multicolumn{2}{c}{Mg I} & \multicolumn{2}{c}{Mg II} & \multicolumn{2}{c}{Ca II} & \multicolumn{2}{c}{Fe II} & \multicolumn{2}{c}{Co II} & \multicolumn{2}{c}{Ni II} \\
&
&
&
&
$\tau$ &
$v_e$ &
$\tau$ & 
$v_e$ & 
$\tau$ & 
$v_e$ & 
$\tau$ & 
$v_e$ & 
$\tau$ &
$v_e$ &
$\tau$ &
$v_e$ & 
$\tau$ &
$v_e$\\
\hline\hline
 Mar 06(+33.4) & 9 & 14 & 10 & 2$\pm$0.2 & 2 & 3$\pm$0.3 & 2 & 0 & 2 & 130$\pm$30 & 3 & 6$\pm$1.5 & 2 & 15$\pm$4 & 3 & 10$\pm$3.5 & 2 \\
 Mar 07(+34.4) & 9 & 14 & 10 & 2$\pm$0.2 & 2 & 3$\pm$0.3 & 2 & 0 & 2 & 100$\pm$25 & 3 & 6$\pm$1.5 & 2 & 15$\pm$4 & 3 & 10$\pm$3.5 & 2 \\
 Mar 12(+39.1) & 9 & 8 & 10 & 2$\pm$0.2 & 2 & 0 & 2 & 0 & 2 & 100$\pm$25 & 3 & 9$\pm$2 & 2 & 15$\pm$4 & 3 & 10$\pm$3.5 & 2 \\
 Mar 17(+44.4) & 9 & 8 & 10 & 3$\pm$0.3 & 2 & 0 & 2 & 0 & 2 & 100$\pm$25 & 3 & 15$\pm$3.5 & 2 & 15$\pm$4 & 2 & 10$\pm$3.5 & 2 \\
 Apr 08(+66.2) & 8 & 8 & 06 & 3$\pm$0.3 & 2 & 0 & 2 & 0 & 2 & 250$\pm$65 & 3 & 60$\pm$15 & 2 & 15$\pm$4 & 2 & 15$\pm$5 & 2 \\
 May 08(+96.2) & 6 & 8 & 06 & 30$\pm$4 & 2 & 40$\pm$5 & 2 & 30$\pm$7 & 2 & 900$\pm$300 & 3 & 50$\pm$13 & 2 & 15$\pm$4 & 2 & 0 & 2 \\
 May 09(+97.2) & 6 & 8 & 06 & 60$\pm$8 & 2 & 40$\pm$5 & 2 & --- & --- & 900$\pm$300 & 3 & 40$\pm$10 & 2 & 35$\pm$10 & 2 & 0 & 2 \\
 May 11(+99.1) & 6 & 6 & 06 & 30$\pm$4 & 2 & 35$\pm$5 & 2 & --- & --- & 20$\pm$4 & 3 & 15$\pm$3.5 & 2 & 20$\pm$5 & 2 & 0 & 2 \\
\hline\hline
\end{tabular}
\caption{SYNOW parameters for Late spectra models. \label{tab:latemaxparams}}
\end{table*}

\begin{figure*}
\centering
\includegraphics[scale=2.0]{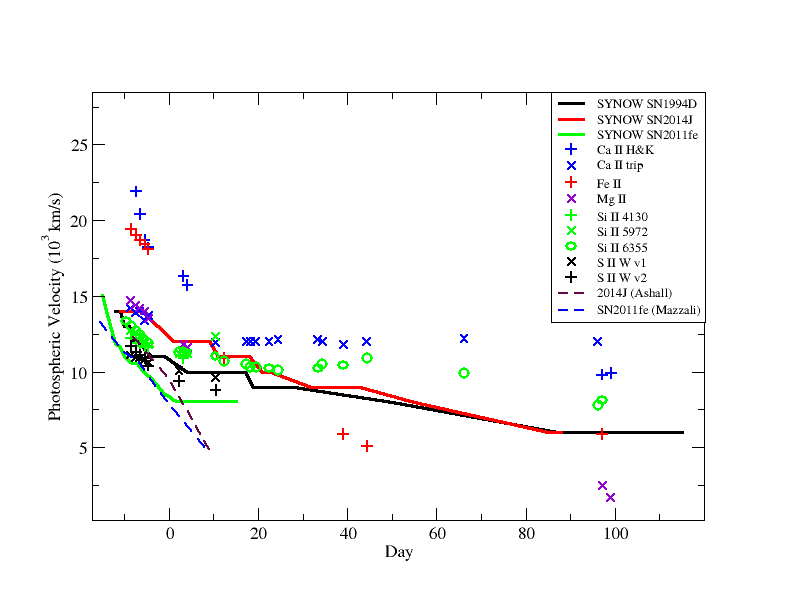}
\caption{\label{fig:vphot}The photospheric
  velocities of three supernovae: SN~1994D \citep{branchcomp105}, SN~2014J 
 \citep[this work,Paper I][]{ashall14J14}, and SN~2011fe
\citep{jerod11fe12,mazz11fe14} versus time.  The SYNOW model velocity for SN~1994D
\citep{branchcomp105} is shown in black, 
the SYNOW model velocity for SN~2014J (this work) is shown in red, SN~2014J \citep{ashall14J14} is
shown in maroon, the SYNOW model velocity for SN~2011fe \citep{jerod11fe12} is shown in
green, and SN~2011fe \citep{mazz11fe14} is shown in blue. The points show the observed line velocities taken from Table~2 of \citet{galbany14J15}.}
\end{figure*}

\section{Photospheric Velocity Comparisons}
\label{sec:vphotcompare}

Figure~\ref{fig:vphot} shows the photospheric velocities of three
supernovae: SN~1994D \citep{branchcomp105}, SN~2014J
\citep[this work; Paper I;][]{ashall14J14}, and SN~2011fe
\citep{jerod11fe12,mazz11fe14} versus time.  The
photospheric velocities of SN~2014J and SN~1994D decrease at a similar
rate, although SN~2014J maintains a photospheric velocity about 1,000
\kmps faster than SN~1994D. The photospheric velocity of SN2011fe, on
the other hand, decreases at a much faster rate, dropping from 15,000
\kmps to 8,000 \kmps in only 14 days. 
The velocity evolution of abundance tomography of SN 2011fe is similar
to that found with SYNOW
\citep{mazz11fe14,jerod11fe12}, but that of SN~2014J is somewhat faster than found
with SYNOW \citep{ashall14J14}. To a large extent the photospheric velocity and the density
slope are degenerate so a shallow density slope can be traded off
for a lower photospheric velocity. The slope of the P-Cygni feature
from peak to tail can be used to break this degeneracy, but only if
one is sure that blending is not important \citep{b93j3,bhm96}. 

Figure~\ref{fig:vphot} also shows the velocities of the ions obtained
in Paper I \citep{galbany14J15}. The results are in good
agreement. Paper I finds high velocity features of Fe II and Ca H+K at
early times  with Si II $\lambda 6355$,  $\lambda 5972$, $\lambda
4130$ match the 
photosphere well (being a little bit slower than inferred by SYNOW, until
the photosphere recedes below the region of primary silicon
formation. The trend is similar for the S~II lines, although they were
measured to be marginally slower than the
Si~II lines. The Mg~II lines and the Ca IR triplet
lines match the early photosphere very well, until hanging up as the
photosphere recedes.   The
evolution of Fe~II lines is that  at early epochs that show a high velocity
component, match well the inferred SYNOW photosphere at about +20 days, fall significantly below the inferred SYNOW photosphere
in the +40-50 Day epochs, and then rise again to roughly match the
inferred photosphere at Day +100. The variation gives an estimate of
the error in inferring ion velocities purely observationally and via a
SYNOW approach. While we find a constant photosphere velocity at early
times, Paper I finds that the strong lines such as Si II $\lambda 6355$
decrease with time. This is not in contradiction, since strong
resonance lines can extend to high velocities at early times, when the
density is high in the outer ejecta.

\section{Comparison with SN 2011\lowercase{fe}}
\label{sec:compare}

SN~2011fe is also a nearby and well studied Type Ia supernova. Here,
we briefly discuss to what extent a comparison between SNe~2014J
and SN~2011fe properties allow us to infer properties about their
progenitors and explosion mechanism, although a complete study is
beyond the scope of this paper.  SN~2014J and SN~2011fe have similar
maximum magnitudes, for example, in $B$  SN~2014J: $-19.19 \pm 0.1$
\citet{marion14J15}, $-19.26 \pm 0.1$ \citet{Kawabata14J14}; 2011fe:
$-19.21 \pm 0.1$ \citet{RS11fe12}, $-19.45 \pm 0.08$
\citep{TR13}, and postmaximum decline rates $\Delta m_{15}$,
SN~2014J: $1.06 \pm 0.06$, averaging the reported values
\citep{tsvetkov14J14,ashall14J14,Kawabata14J14,marion14J15}; SN~2011fe:
$1.11 \pm 0.07$ \citep{TR13,RS11fe12,mcclelland11fe13,factory11fe13} and different
line and photospheric velocities, up to 
1500 \kmps around maximum, as shown in Fig~\ref{fig:vphot}.
Neither SN~2011fe nor SN~2014J
show traces of a single degenerate companion
(\citealp{lundqvist14J15}, but see
also \citealp{Graham14J15}) favoring, in principle, the 
double degenerate scenario \citep[see also][]{soker11fe14,soker14J15}.  SN~2014J is highly reddened E(B-V) = 1.2
\citep{Goobar14j14}, while SN~2011fe is not, E(B-V) = 0.03
\citep{mazz11fe14}.  Observed properties of both supernovae can be
explained by the delayed detonation explosion of a carbon-oxygen white
dwarf (WD) with a mass close to the Chandrasekhar mass. In delayed
detonation explosion models \citep{khok91a,hkw95,hk96}, the main
properties for the outcome are \citep{hoeflich10}: the
composition of the WD --- related to the WD progenitors; the central
explosion ignition density --- related with the accretion rate and/or,
in the case of rotation, the time scale of angular momentum loss; and
the transition density from deflagration to detonation --- the
main factor that determines the \nni mass, which is a free
parameter. To reproduce the light curves, \nni masses above 0.51-0.55
and 0.6-0.65~\msol for SNe~2011fe and SN~2014J, respectively, have been
estimated \citep{Isern11fe13,b11fe15}.  Higher transition densities
(lower pre-expansions) will produce more \nni and less intermediate
mass elements, but the corresponding increase in the kinetic energy
would be above a factor of 5 smaller than the increase in \nni masses.
The
difference in ejecta velocities between SNe 2014J and 2011fe is
$\Delta v_{ej}
\sim 1500$~\kmps, and the ejecta velocity is $v_{ej}\sim 10000$~\kmps,
and so the 
difference is $\sim 15$\%.  The difference in \nni mass between SNe 2014J and
2011fe is $\Delta \mni\sim 0.1$, and the nickel mass is $\mni \sim
0.6$, and so the 
difference is $\sim 17$\%, therefore the difference in
ejecta velocity should be more like 3\% if a variation in the
transition density alone were the 
underlying physical difference between SNe 2014J and 2011fe.
In fact, this is
the reason why delayed-detonation models reproduce the Phillips
relation: when the transition density  is modified, different \nni
masses are obtained 
for similar kinetic energy, produced ad hoc, with pre-deflagrations,
or via other models. See for example \citet{hk96}, (their Table
1), for a delayed-detonation model denoted M36 that produces 0.6\msol of
\nni. A smaller transition density (model M37) changes \nni by 16\%
and $E_{kin}$ by only 2\%.

This implies that the differences in the ejecta velocities of SN~2014J
with respect to SN~2011fe can not be explained by just increasing the
transition density. The observed variation in velocities can possibly, be
explained assuming different central ignition densities or different
WD chemical compositions, mainly the ratio C/O \citep{DHS01b}.

Smaller central ignition densities produce more \nni and less
neutronized elements in the central regions and, for smaller central
densities, the WD binding energy is also smaller and so, the kinetic
energy is higher, while maximum magnitudes and decline rates could be
similar for the density range needed to explain these two supernovae.
Thus, SN~2014J could have been ignited at a central density smaller
than that for SN~2011fe, due to a higher accretion rate or, in case of
rotation, to a shorter time scale for the angular momentum losses
\citep{Piersanti03a,dominquez06}.  If this were the case, the light
curve tail of SN~2011fe would be dimmer than that of SN~2014J, as it
seems to be in $B$ band and less markedly in $V$ \citep{Foley14J14},
although late observations in the optical of 
SN~2014J have not yet been published.  A greater C/O ratio will produce
more \nni and more kinetic energy. In fact, an increase of C/O above 
20\% will produce an increase above 10\% in \nni mass, an increase
above 1500~\kmps in line velocities and 2000~\kmps in the photospheric
velocity \citep{DHS01b}.  If the difference is due to the differences
in the C/O ratio between the progenitors of SN~2014J and SN~2011fe,
their light curve tails will be similar --- for the range in C/O needed
to explain the observed velocities --- and SN~2014J will be 0.1
magnitudes brighter at maximum light.  
Tail
brightness (we mean here more than 40 days after maximum)
reflects the total \nni mass, for SN 2011fe we interpret the slightly
dimmer tail, as less \nni mass ($\sim 0.1$~\msol). At the same time the
overall velocities are smaller, while the maximum magnitude and
decline rate $\Delta m_{15}$ are similar. For a higher central
density, the kinetic energy 
is smaller due to the higher binding energies and electron captures
are favored, producing neutronised elements at the center. Thus, less
\nni is produced (in agreement with the light curve 35--40 days after
maximum), but 
this \nni is located closer to the surface, so at maximum the \nni
decrease is compensated by the its more external location. Second, for
these two SNe we know the distance. In
SNe Ia at unknown distance we should work with the ``peak to tail
ratio'' and then the uncertainty in the maximum magnitude would
be critical \citep[see][]{hoeflich10}.
Reported observations show that
SN~2014J is dimmer at maximum than SN~2011fe, but note that 0.1 mag is
within the errors, SN~2014J is heavily reddened and there are
discrepancies in the maximum magnitude as derived by different
authors. The chemical composition of the exploding WD is related to
the properties of the WD progenitors, mainly the initial main sequence
mass and, to a smaller extent, the initial metallicity.

Assuming that both WDs are produced in the double degenerate scenario, a decrease of 
in C/O above 20\% are expected just by changing the the initial accreting
WD mass from 0.8 to 0.9 \msol, corresponding to the C+O cores of main
sequence masses 5 and 6 \msol (initial solar composition) and up to
35\%, if a 1~\msol WD (initial main sequence mass 7~\msol) is considered \citep{bravo_met10}.
Smaller differences, in terms of \nni masses, would be
obtained assuming different WD cooling times, diffusion and
crystallization processes, before accretion, for example, an increase of 7\% in \nni mass increasing 
 the cooling time from 0.6 to
0.8 Gyr, \citep{bravo_chem11} or changing the metallicity \citep{bravo_met10}, i.e. from
solar to sub-solar metallicities, as suggested for SN~2011fe
\citep{mazz11fe14,b11fe15}.  In both
cases, the reason is the central abundance of $^{22}$Ne that changes the
neutronization and so, the nucleosynthesis (less $^{22}$Ne, less
neutronization and more \nni), but the corresponding changes in kinetic
energy/velocities would be smaller than in the previous
cases, that is due to varying the C/O ratio  or the central density.
We note the the C/O ratio was studied in 3-D in the context of pure
deflagration models by \citet{RH04a} who did not find a significant
variation in the total mass of \nni produced in their simulations with
their particular burning prescription. Further work by this same group
extending to the case of delayed detonation models; however, found
complete agreement with the work of \citet{DHS01a} and
\citet{hoeflich10}, that is they found the above effects of the C/O
ratio on the mass of \nni produced.

In summary, it may be that SN~2014J comes from a smaller primary (and
corresponding main sequence) WD mass compared to SN~2011fe, explaining the
different observed velocities at maximum, for similar decline
rates. In this case similar light curve tail brightnesses would be
expected for the range of C/O needed.  If the observed maximum magnitude for both SNe were correct
and SN~2014J is slightly dimmer at maximum, but showing a brighter light
curve tail than SN~2011fe (to be confirmed), it is more likely that
SN~2011fe had a greater central ignition density than SN~2014J, this would
also explain the smaller expansion velocities of 2011fe. 
For SN~2011fe a sub-solar
metallicity, about $Z_\odot$/20, has been suggested \citep{mazz11fe14,b11fe15}.
If SN~2011fe has a lower progenitor metalicity than that of SN~2014J,
SN~2011fe would be expected to be brighter than SN~2014J, everything
else being equal. However, this would slightly increase its
maximum magnitude. The metalicity difference between SN~2011fe and SN
2014J is likely small, therefore it would not cause significant
variations in the brightness at maximum light.

\section{Discussion and Conclusions}
\label{sec:conclusion}

\citet{ashall14J14} performed a different type of spectral
analysis on SN 2014J and it is worthwhile to compare the results of the two
methods. We are in fact mostly in agreement with the conclusions of
\citet{ashall14J14}, but we differ in some cases and we discuss where we
agree and disagree.
We do not calculate abundances to compare with \citet{ashall14J14},
since in SYNOW the excitation temperature is just a modeling
parameter. 
While the early spectra
show clear evidence for standard SNe~Ia features, we do not find
carbon in the spectra, nor do we find oxygen, both of which were found
by \citet{ashall14J14} at early times. We also do not find 
strong evidence for 
mixing of nickel to high velocities. We find weak evidence for Si~III
post maximum (up to Day 23) and \citet{ashall14J14} find no evidence
for Si~III after Day -7 (Jan 27). The evidence for Si~III is not very
strong, though, so this
difference is not significant.  
The general results from INTEGRAL, which detected the 847  and 1238 keV
$\gamma$-ray 
lines of radioactive $^{56}$Co
\citep{churazov14J14,churazov14J15,Diehl14J14,Diehl14J15}, showed that
the general nickel distribution was stratified with a total \nni mass
of $\sim 0.5$~\msol. Due
to the low signal to noise there is some ambiguity about just how to
interpret the results, that is, whether the $\gamma-$rays favor
symmetric nickel distributions \citep{churazov14J14,churazov14J15} or
whether the lines are significantly Doppler broadened 
asymmetric ones \citep{Diehl14J14,Diehl14J15}. 
We are not terribly sensitive to small asymmetries, and therefore we
cannot distinguish between these different interpretations. However,
our results are in general agreement of with the underlying \nni
distribution inferred from the INTEGRAL data.

\begin{figure}
\centering
\includegraphics[scale=0.55]{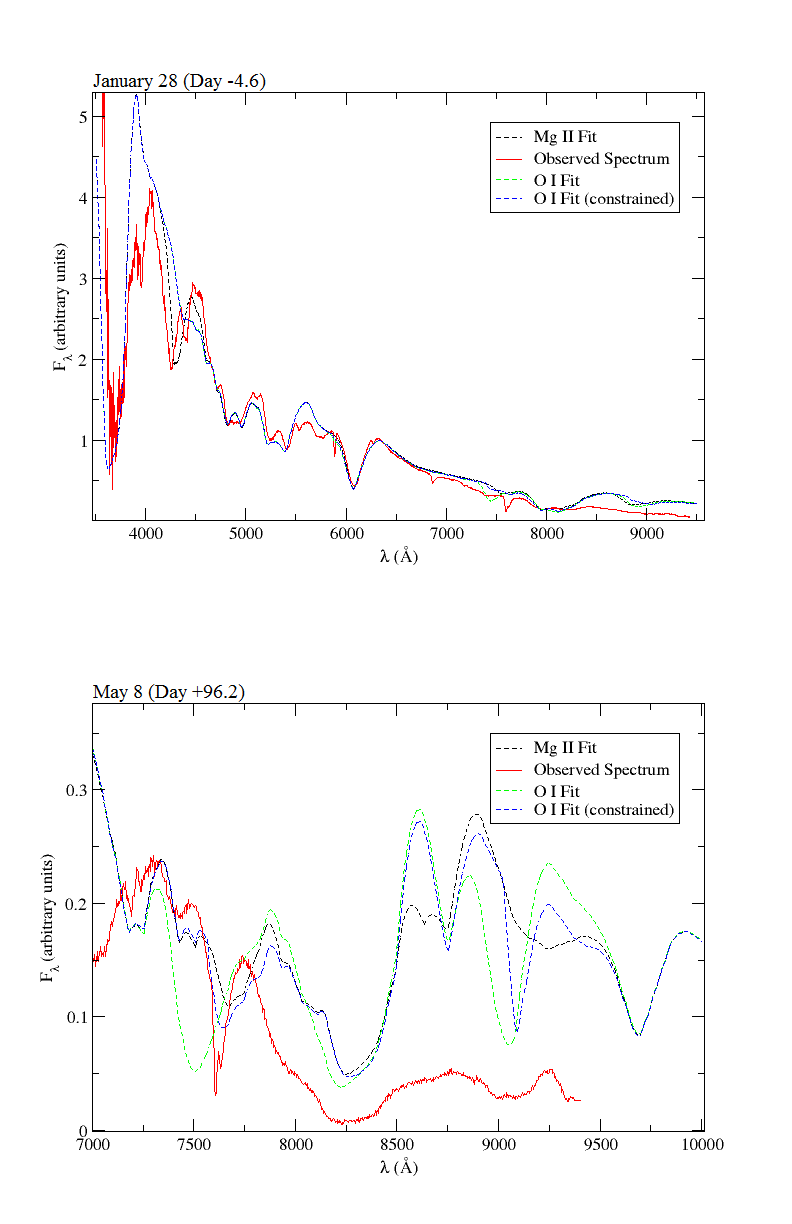}
\caption{\label{fig:MgO}Comparison between
  Mg II and O I in the synthetic spectra. The upper panel shows an early spectrum (Day -4.6), which emphasizes the need for Mg II to model
  the absorption feature near 4300 {\AA}. The lower panel shows a late
  spectrum (Day +96.2). While Mg II models the 7600 feature decently well, the O I
  absorption in this region is far too blue unless oxygen is
  constrained to a maximum velocity a mere 1,000 km/s above the
  photospheric velocity. It is worth noting, though, that the O I absorption 
  near 9100 {\AA} does seem to model a feature of the observed spectrum, so we acknowledge
  the possible presence of oxygen at late times.}
\end{figure}

While abundance tomography \citep{ashall14J14} indicates the presence
of significant 
oxygen, at early times, we do not find any spectroscopic
signatures for oxygen 
at early times and can only weakly identify it at late times.
Mg II models the P-Cygni profile around 4350 {\AA} in the early and
the near-maximum spectra, while O~I does not. While Mg II doesn't have the
same strong identifying feature at 4350 {\AA} in the post-maximum spectra,
it does model the 7600 feature decently well, while adding O~I
causes an absorption a hundred or so angstroms too blue unless O~I is
constrained to be within $1,000$~\kmps of the photospheric
velocity. This is illustrated in Figure~\ref{fig:MgO}. It seems odd
that O~I would be confined to a region so close to the photosphere
even though the photosphere has moved from 14,000~\kmps to
6,000~\kmps.

There is some reason to doubt the Mg~II identification, though,
as post maximum spectra do not show
a strong feature at Mg II $\lambda\lambda9226$, and the O~I feature does
seem to correspond the absorption at 9100 {\AA} somewhat. As such, and
considering SYNOW's limited applicability at late times, we can't
confidently rule out the presence of oxygen at late times. Also, the
O~I identification is a bit suspect since the O~I 7774 line is modeled
well by the constrained ion, the 9100~\AA\ feature is perhaps better
modeled by the unconstrained ion. Thus, we should be careful
about drawing 
conclusions about mixing of unprocessed or only slightly processed
elements to low velocities. 

Just because there is little spectroscopic evidence for oxygen does not
mean that it is not there. Since O~I has an ionization potential for
13.6~eV, it requires higher temperatures to populate the excited
states and so at low temperatures oxygen can be spectroscopically
hidden. However, we also do not find any evidence for mixing of \nni
to high velocities. One would expect that if radioactive \nni is mixed
to high velocities, then non thermal excitation of oxygen from fast
electrons created by $\gamma$-ray interactions should be available to
excite and ionize significant amounts of oxygen, which leads to strong
O~I lines. We assume the spectroscopic rule of thumb that the strongest
lines of most ions come usually when it is not the dominant ion, but
rather when the next ionization state is dominant. This is true
because then there is plenty of energy to populate the excited states
of the lower ionization stage.

Within the limits of our analysis we show that there is not
significant oxygen in the observed spectra and that there appears to
be no need for significant mixing of \nni into the outer layers of
SN~2014J. We find good agreement with the photospheric velocity from
other groups using a similar approach, particularly with Paper I. The fast decline found in the
photospheric velocity by \citet{ashall14J14} is likely due to a
difference in definition of the photosphere, due to their  Monte-Carlo approach rather
than some important systematic difference in the methods, but
the difference in definition is important to keep in mind when
comparing velocities determined by different methods. The
  evidence that the definitions are different can be seen in
  Fig~\ref{fig:vphot} where the trend in velocities of Si II $\lambda 6355$
  found by \citet{ashall14J14}
  follow those that were strictly measured from the spectra in Paper I.

The late time spectra shown in Figure~\ref{fig:late}  are in good
general agreement with the more detailed modeling results in the IR
\citep{brian_14JIR14} and in the Mid-IR
\citep{telesco14J15}. In general we find features dominated by iron
group elements, 
with features from intermediate group elements such as Ca~II and Mg~II
whose lines become strong at low temperatures. That is, we find
evidence for a layered structure that is generically predicted by
delayed detonation and pulsating delayed detonation models.

In conclusion, our SYNOW modeling indicates that 2014J is a relatively normal SN Ia.
Si II, S II, Mg II, and Ca II features dominate the spectra at early times.
These ions are joined by Na I and Mg I as we approach maximum light, and at later times iron-group elements also appear, as expected in the stratified abundance model of the formation of normal type Ia SNe.
We generally agree with the findings of \cite{ashall14J14}, but while
abundance tomography indicates the presence of significant oxygen
at early epochs, we
do not find strong evidence to support this, as we find no
spectroscopic signatures for oxygen at early times and only very weak
signatures at late times. By strong evidence, we see no
  evidence for the need for the O I 7774 line, despite the fact that
  there is a telluric feature. In normal SNe~Ia, the O I 7774 is quite
evident even with the telluric feature \citep[see][]{branchcomp105}. Using our results and other published
results we are able to account for the similarities and differences
between SN 2014J and SN 2011fe.

\section*{Acknowledgments}
This work is based on service observations (program
SW2013a08, P.I. P- Ruiz-Lapuente)
made with the 
William Herschel Telescope 
(WHT), and on discretionary observations made with the Isaac Newton
Telescope (INT), both operated on the island of La Palma by the Isaac Newton
Group (ING) in the Spanish Observatorio del Roque de los Muchachos of the
Instituto de Astrof\'isica de Canarias. We thank the ING Director for having
made public the INT data as soon as they were obtained. We also acknowledge
the observers who kindly donated their time to monitor SN2014J on both the
WHT and the INT. We thank Mercedes Molla for the proposal of 
observation time. 
The work has been supported in part by
support for program
HST-GO-12948.004-A   provided by NASA through a grant from the
Space Telescope Science Institute, which is operated by the
Association of Universities for Research in Astronomy, Incorporated,
under NASA contract NAS5-26555.
EB acknowledges support from a Carl Bush Fellowship from the
University of Oklahoma.
Support for LG is provided by the Ministry of Economy, Development,
and Tourism's Millennium Science Initiative through grant IC120009,
awarded to The Millennium Institute of Astrophysics, MAS. LG
acknowledges support by CONICYT through FONDECYT grant 3140566. 
JIGH acknowledges financial support from the Spanish Ministry of 
Economy and Competitiveness (MINECO) under the 2011 Severo Ochoa 
program MINECO SEV-2011-0187, and the 2013 Ram\'on y Cajal program 
MINECO RYC-2013-14875, and the Spanish ministry project MINECO 
AYA2014-56359-P. ID 
acknowledges MINECO-FEDER grant AYA2011-22460.

\bibliography{apj-jour,mystrings,refs,baron,sn1bc,sn1a,sn87a,snii,stars,rte,cosmology,gals,agn,atomdata,crossrefs}

\label{lastpage}

\end{document}